\documentclass[twocolumn,prb,color,psfig,superscriptaddress]{revtex4}
\usepackage{graphicx}
\usepackage{epsfig}

\def\beq{\begin{equation}}
\def\eeq{\end{equation}}

\begin{document}

%\title{Non-Zhang-Rice states and unconventional order parameters for hole CuO$_4$ centers in cuprates}
\title{Pseudogap phase in cuprates: oxygen orbital moments instead of circulating currents.}
%\title{Electron structure of hole centers in CuO$_2$ planes of cuprates }
%\title{Non-Zhang-Rice structure of hole centers in CuO$_2$ planes of cuprates }
%\title{Interplay between bonding $\sigma$ and nonbonding $\pi$ oxygen hole orbitals and non-Zhang-Rice states in CuO$_2$ planes of cuprates }
\author{A.S.~Moskvin}
\affiliation{Ural Federal University, 620083 Ekaterinburg,  Russia}
\date{\today}

\begin{abstract}
Circulating current (CC) loops within the cuprate unit cell are proposed to play a key role in the physics of the pseudogap phase.  
However,  main experimental observations motivated by this sophisticated proposal  and seemingly supporting the CC model can be explained in frames of a simple and physically clear microscopic model. We argue that instead of a well-isolated Zhang-Rice (ZR) singlet $^1A_{1g}$ the ground state of the hole center [CuO$_4$]$^{5-}$ (cluster analog of Cu$^{3+}$ ion) in cuprates should be described by a complex $^1A_{1g}$-$^{1,3}B_{2g}$-$^{1,3}E_u$ multiplet, formed by a competition of conventional hybrid Cu 3d-O 2p $b_{1g}(\sigma)\propto d_{x^2 -y^2}$ state and {\it purely oxygen nonbonding} O 2p$\pi$ states with $a_{2g}(\pi)$ and $e_{ux,y}(\pi)$ symmetry. In contrast with inactive ZR singlet we arrive at several novel competing orbital and spin-orbital order parameters, e.g.,  Ising-like net orbital magnetic moment, orbital toroidal moment, intra-plaquette's staggered order of Ising-like oxygen orbital magnetic moments. As a most impressive validation of the non-ZR model we explain fascinating results of recent neutron scattering measurements that revealed novel type of magnetic ordering in pseudogap phase of several hole-doped cuprates.

%The problem of the order parameters in cuprates is closely related to that of ground state for the hole centers.

%Note, Varma and collaborators [83] argue that this broken symmetry order plays a key role in the physics of the pseudogap phase.

%There have been, however, a few concise proposals which make falsifiable predictions. Intellectual masterpieces among them have been the theory of  a more recent proposal that the anomalous properties of the cuprates may be due to quantum critical fluctuations of current patterns formed spontaneously in the CuO planes.
 
\end{abstract}

\maketitle
{\bf Introduction}.
The nature of the pseudogap (PG) phase to be the most puzzling and anomalous region of
the phase diagram of the cuprate high-temperature superconductors is one of  major unsolved problems in condensed matter physics. Different theoretical models describe the pseudogap state
as a precursor of the superconducting d-wave gap  with preformed pairs below T$^*$ which would acquire phase coherence below T$_c$ or
 a phase competing with the superconducting one  that ends at a quantum critical point, typically inside the superconducting dome.

The order parameter associated with these competing phases may involve charge and spin density waves or charge currents flowing around the CuO$_2$ square lattice, such as D-charge density wave  or orbital circulating currents (see, e.g., Ref.\onlinecite{Rice} for a short overview).
In his theory
for cuprates, C. M. Varma\,\cite{Varma} proposes that PG is a
new state of matter associated with the spontaneous appearance of circulating current (CC) loops within CuO$_2$ unit cell. The current pattern is assumed to disappear only at a  quantum critical point at a hole doping level of $x_c$\,$\sim$\,0.19. 
 This Intra-Unit-Cell (IUC) order breaks time reversal symmetry (TRS), but preserves lattice translation invariance. 

From a theoretical point of view the existence of a CC-loop order and the ability of such a q=0
instability to produce a gap in the charge excitation spectrum are still highly controversial\,\cite{anti-Varma,Weber}.  However, several experimental observations motivated by this proposal pointed to a symmetry breaking in the pseudo-gap phase\,\cite{ARPES,Bourges,Kerr,Bi2212,three} and provided strong encouragement for models based on CC-loop order in copper oxide materials.
The TRS violation in the PG state of Bi2212 was first inferred from the observation of dichroic effect in ARPES measurements\,\cite{ARPES}, but this measurement has been the subject to a long standing controversy (see, e.g., Ref.\,\onlinecite{anti-ARPES}).

%There is now a growing number of experimental indications that the pseudo-gap phase actually corresponds to a symmetry breaking state [1–5]. 

%Motivated by this proposal, several experimental groups have looked for signatures of orbital currents or T violation in CuO superconductors. While there is no agreement between different groups regarding the manifestation of T violation in angle-resolved photoelectron spectroscopy studies,36,37 a recent neutron scattering experiment by Fauqué et al.12indicates magnetic order within the unit cells of the CuO planes.

Seemingly the strongest experimental evidence for an orbital-current phase are the observations of an unusual translational-symmetry preserving magnetic order in underdoped YBa$_2$Cu$_3$O$_{6+x}$,  HgBa$_2$CuO$_{4+\delta}$, La$_{2-x}$Sr$_x$CuO$_4$, and Bi$_2$Sr$_2$CaCu$_2$O$_{8+\delta}$ by spin-polarized neutron diffraction\,\cite{Bourges,Bi2212}. However, local probes of magnetism, such as nuclear magnetic
resonance (NMR) and zero-field muon spin relaxation (ZF-$\mu$SR) have found no evidence for the onset of magnetic order at the pseudogap temperature T$^*$\,\cite{17O-NMR,Strassle,Sonier}. Very recent high-precision ZF-$\mu$SR measurements of La$_{2-x}$Sr$_x$CuO$_4$ in the
Sr concentration (hole-doping) range 0.13\,$<$\,x\,$<$\,0.19\,\cite{Sonier} do not
support theoretically predicted loop-current phases\,\cite{Varma}, and point to
 an alternative explanation for the unusual magnetic order
detected by spin-polarized neutron diffraction at lower hole
doping. Hereafter in this Letter we address such an alternative scenario.

{\bf Non-Zhang-Rice model and order parameters}.
The problem of the order parameters in hole doped cuprates is closely related to that of ground state of the hole centers CuO$_4^{5-}$ to be  cluster analogues of the Cu$^{3+}$ ion. 
%The pseudogap phase in high Tc superconductors has attracted intense research efforts in recent years. This phase is expected to be intimately linked with the state of an excess O hole doped in the CuO2 plane. 
The nature of the doped-hole state in the cuprates with nominally Cu$^{2+}$ ions such as La$_2$CuO$_4$ is a matter of great importance in understanding both the mechanism leading to the high-temperature superconductivity and unconventional normal state behavior of the cuprates. In 1988 Zhang and Rice\,\cite{ZR} have proposed that the doped hole forms a well isolated local spin and orbital ${}^1A_{1g}$ singlet state which involves a phase coherent combination of the 2p$\sigma$ orbitals of the four nearest neighbor oxygens with the same $b_{1g}$ symmetry as for a bare Cu 3$d_{x^2-y^2}$ hole. 
This all assumes that, in the low energy limit, the Cu-O system can be reduced to
an effective single orbital, or one-band model.

% However this simplification was called into question, and it was pointed out that it is necessary to retain the full three band nature of the model in order to capture the important physics.

However, numerous experimental data, in particular, recent magnetic neutron scattering findings (see review article Ref.\,\onlinecite{Bourges}), suggests the involvement of some other physics  which introduces low-lying states into the excitation of the doped-hole state, or competition of conventional Zhang-Rice (ZR) state with another electron removal state.  This point was discussed earlier, however, mainly as an interplay between ZR singlet ${}^{1}A_{1g}$ and triplet ${}^{3}B_{1g}$, formed by additional hole going not into $b_{1g}$ state as in ZR singlet, but into $a_{1g}\propto d_{z^2}$ state\,\cite{Eskes}. It is worth noting that ${}^{3}B_{1g}$ state corresponds to a Hund ${}^{3}A_{2g}$ term of two-hole $e_g^2$ configuration of an undistorted CuO$_6$ octahedra. However, later experimental findings for very different insulating cuprates  and theoretical calculations have shown that the energy separation between  the $b_{1g}$(d$_{x^2-y^2}$) and $a_{1g}$(d$_{z^2}$) orbitals in CuO$_4$ plaquettes is thought to be of the order of 1.5\,eV, i.e. too large for quasi-degeneracy and effective  vibronic coupling. More sophisticated version of the non-ZR states was proposed by Varma\,\cite{Varma}, who has proposed that the additional holes doped in the CuO$_2$ planes do not hybridize into ZR singlets, but give rise to  circulating currents on O-Cu-O triangles.

On the other hand,  cluster model considerations supported by numerous experimental data point to a competition of conventional hybrid Cu 3d-O 2p $b_{1g}(\sigma)\propto d_{x^2 -y^2}$ state with {\it purely oxygen nonbonding} O 2p$\pi$ states with $a_{2g}(\pi)$ and $e_{ux,y}(\pi) \propto p_{x,y}$ symmetry (see Refs.\onlinecite{NQR-NMR,Moskvin-FNT-11} and references therein). Accordingly, the ground state of such a non-ZR hole CuO$_4^{5-}$ center  as a cluster analog of Cu$^{3+}$ ion should be described by a complex $^1A_{1g}$-$^{1,3}B_{2g}$-$^{1,3}E_u$ valence multiplet with several order parameters such as  spin and Ising-like orbital magnetic moments, dipole and quadrupole electric moments prone to strong vibronic coupling, and more subtle hidden order parameters.

Despite a large body of both theoretical and experimental argumentation indirectly supporting the existence of non-ZR multiplets in cuprates their direct experimental probing remains to be highly desirable especially because there are  numerous misleading reports supporting "the stability of simple ZR singlet". For instance,  the authors of the 
photoemission studies on CuO and Bi$_2$Sr$_2$CaCu$_2$O$_{8-\delta}$\,\cite{Tjeng}, have reported that they "are able to unravel the different spin states in the single-particle excitation spectrum of cuprates and show that the top of the valence band is of
pure singlet character, which provides strong support for the existence and
stability of Zhang-Rice singlets in high-$T_c$ cuprates thus
justifying the ansatz of single-band models".  In their opinion "these states are more stable than the triplet states by about 1\,eV".  However, in their
photoemission studies they made use of the Cu 2p$_{3/2}(L_{3})$ resonance
condition that allows to detect unambiguously only copper photo-hole states,
hence they cannot see the purely oxygen photo-hole $a_{2g}$ and $e_u$ states. 
  
\begin{figure}[t]
\includegraphics[width=6.5cm,angle=0]{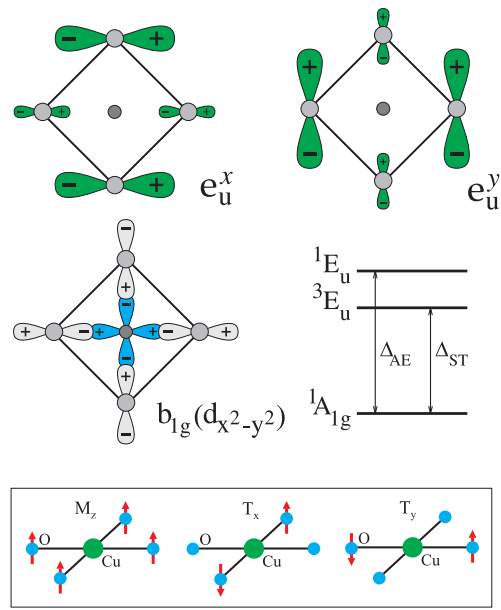}
\caption{(Color online) The term  structure of the actual  valence A-E multiplet for a hole
CuO$_{4}^{5-}$ center together with the single-hole $b_{1g}^{b}$ and
$e_{ux,y}^{b}$  orbitals. Lower panel illustrates the ferromagnetic and toroidal orderings of the oxygen orbital magnetic moments within the CuO$_4$ plaquette.} 
\label{fig1}
\end{figure}

Earlier we have addressed unconventional properties of the non-ZR hole center related to the $^1A_{1g}$-$^{1,3}E_u$ quasi-degeneracy (A-E model)\,\cite{NQR-NMR}. Fig.\,\ref{fig1} shows the term  structure of the actual  valence A-E multiplet together with single-hole basis $b_{1g}^{b}$ ($|b_{1g}\rangle =c_d |d_{x^2 -y^2}\rangle +c_p|b_{1g}(O2p)\rangle $) and $e_{ux,y}$ ($|e_{ux,y}\rangle =c_{\pi} |e_{ux,y}(\pi)\rangle +c_{\sigma}|e_{ux,y}(\sigma)\rangle $) orbitals. 
The $e_{u}$ orbitals could form two  circular current $p_{\pm 1}$-like states, $e_{u\pm 1}$
  with an Ising-like orbital moment $\langle e_{u\pm 1}|l_z|e_{u\pm 1}\rangle = \pm 2c_{\sigma}c_{\pi}$ which is easily prone to be quenched by a low-symmetry crystal field with formation of two   currentless, e.g., $p_{x,y}$-like $e_{ux,y}$ states. 

Even with neglecting the spin degree of freedom we arrive at the eight order parameters for the hole [CuO$_4$]$^{5-}$ center including  
both conventional (two-component in-plane electric dipole moment, three-component  in-plane electric quadrupole moment) and unconventional (Ising-like purely oxygen orbital magnetic moment and two-component in-plane purely oxygen orbital toroidal moment) ones (see Fig.\,\ref{fig1}).
 Thus,  the CuO$_4$
plaquette with ($^1A_{1g},^1E_u$) valent multiplet forms an unconventional
magneto-electric center characterized by eight independent orbital order
parameters. Even this simplified  model predicts
 broken  time-reversal ($T$) symmetry, two-dimensional parity ($P$), and basic tetragonal (four-fold Z$_4$) symmetry.
 The situation seems to be more involved, if we take into account spin degree of freedom, in particular, the ${}^1A_{1g}$-${}^{3}E_{u}$ singlet-triplet mixing effects. First of all, such a center is characterized by a true spin S=1 moment being gapped, if the ZR singlet ${}^1A_{1g}$ has the lowest energy. Strictly speaking, for our two hole configuration we should introduce two spin operators: net spin moment ${\hat {\bf S}}={\hat {\bf s}}_1+{\hat {\bf s}}_2$ and  spin operator ${\hat {\bf V}}={\hat {\bf s}}_1-{\hat {\bf s}}_2$ that changes spin multiplicity. It should be noted that the $V$-type order implies an indefinite ground state  spin multiplicity and at variance with $S$-type order is invariably accompanied by an orbital order. The singlet-triplet structure of the A-E multiplet implies two novel types of the spin-orbital order parameters:
spin-dipole parameters $\langle {\hat {\bf V}} {\hat d}_x\rangle$ and $\langle {\hat {\bf V}} {\hat d}_y\rangle$ and spin-toroidal parameters $\langle {\hat {\bf V}} {\hat T}_x\rangle$ and $\langle {\hat {\bf V}} {\hat T}_y\rangle$. Novel ordering does not imply independent V-, d-  or T-type orders. 
%In other words,  both spin-dipole and spin-toroidal orders within CuO$_4$ plaquette are described by  hidden order parameters, that can be revealed only by specific experimental technique, for instance, by magnetic polarized neutron diffraction.

Despite a "fragility" \, of the orbital $e_u$-currents with regard to a crystal field quenching  these can produce ferromagnetic-like fluctuations that can explain numerous manifestations of a weak ferromagnetism in different cuprates (see, e.g., Ref.\onlinecite{WM}) and a remarkable observation of a weak magnetic circular dichroism (MCD) in YBa$_2$Cu$_3$O$_{6+x}$\,\cite{Kerr}.  It should be noted that the value of MCD effect does
 not straightforwardly depends on the value of orbital,
  or magnetic orbital moment.
  Generally speaking, the hole doped cuprate could be a system with a giant circular
  magnetooptics if we were able to realize the uniform ferromagnetic ordering of the
  orbital $e_u$-currents. It seems likely that  the relative concentration $x_{h}\sim 10^{-4}$ of   circularly polarized $e_u$ holes is enough to provide the same magnitude
   of MCD as an applied magnetic field of $1$ Tesla. It is worth noting that the current loop state\,\cite{Varma}, by itself, is incompatible with ferromagnetism and cannot explain the Kerr measurements\,\cite{Kerr}.    

%It should be noted that simultaneously the one-center  CT excitations yield a contribution to the anisotropic symmetric part of the  polarizability for the CuO$_4$ center that determines the in-plane linear birefringeance and  dichroism:

%These expressions yield the so-called $intrinsic$ circular and linear  birefringeance and dichroism, respectively. The magnitude of the macroscopic  observed effects depends in a rather complicated way on the relation between them. Enhanced linear birefringeance in crystals usually leads to an effective suppression of circular effects because the former prefers the linearly polarized light  modes, rather than the circular ones. 
%A simple comparison of Exps.(\ref{g_z}) and (\ref{a1g})-(\ref{b12g}) yields an interesting example of complex interplay between the two effects on the microscopic level. Indeed, the large values of quadrupole moment are typical for $e_{ux,y}$ states with quenched orbital moment, and vice versa.

    %Such fluctuations can easily explain the seemingly contradictory results    in the search for spontaneous gyrotropy in different cuprates\,\cite{Kerr}.  Anycase the topological orbital ordering results in strong light depolarization effects clearly observed, for instance,  in 124 system\,\cite{Wachter}.

It is worth noting that occurrence of  both orbital toroidal and spin-dipole order parameters point to the hole CuO$_4^{5-}$ centers as polar centers with effective magneto-electric coupling which can provide ferroelectric and magnetoelectric properties for hole-doped cuprates\,\cite{ME}. Interestingly, within the $^1A_{1g},^1E_u$ multiplet the electric dipole moment operator can be coupled with orbital toroidal and magnetic moments by a remarkable magnetoelectric relation\,\cite{NQR-NMR,Moskvin-FNT-11}: ${\hat d}_x=d_{me}\{{\hat T_y,{\hat M_z}}\}$, ${\hat d}_y=-d_{me}\{{\hat T_x,{\hat M_z}}\}$. 

\begin{figure}[t]
\includegraphics[width=6.5cm,angle=0]{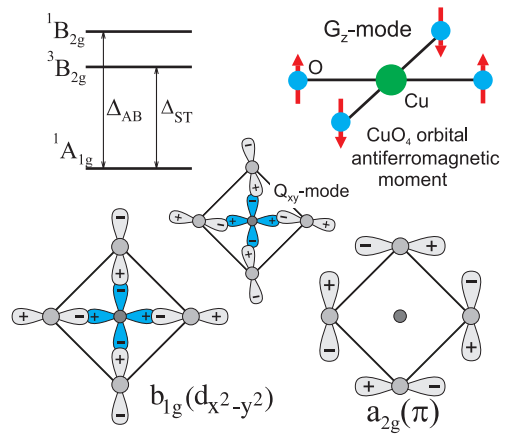}
\caption{(Color online) The term  structure of the actual  valent A-B multiplet for hole
CuO$_{4}^{5-}$ center together with single-hole basis $b_{1g}^{b}$ and
$a_{2g}^{b}$  orbitals. Shown are antiferromagnetic (staggered) ordering of oxygen orbital magnetic moments within CuO$_4$ plaquette ($G_z$-mode) and  quadrupole $Q_{xy}$-mode.} 
\label{fig2}
\end{figure}

Hereafter we address novel effects related with the $^1A_{1g}$\,-\,$^{1,3}B_{2g}$ quasi-degeneracy (A-B-model).    
Unconventional orbital A-B structure of the hole  CuO$_4$ hole centers 
with the ground state $b_{1g}^2$:${}^1A_{1g}$\,-\,$b_{1g}a_{2g}(\pi)$:${}^{1,3}B_{2g}$ multiplet (see Fig.\,\ref{fig2}) implies several spin, charge, and orbital order parameters missed in the simple ZR model. 
For the orbital quasi-doublet ${}^1A_{1g}$\,-\,${}^{1}B_{2g}$ to be properly described one might make use of a pseudo-spin formalism with two states ${}^1A_{1g}$ and ${}^{1}B_{2g}$ attributed to $|+\frac{1}{2}\rangle$ and $|-\frac{1}{2}\rangle$ states of a pseudo-spin $s=\frac{1}{2}$, respectively. Then we introduce three order parameters: $\langle {\hat\sigma}_z\rangle$, $\langle {\hat\sigma}_x\rangle$,  and $\langle {\hat\sigma}_y\rangle$, where ${\hat{\bf \sigma}_i}$ is Pauli matrix. Order parameter $\langle {\hat\sigma}_z\rangle$ defines the symmetry conserving charge density fluctuations within the CuO$_4$ plaquette:
Order parameter $\langle {\hat\sigma}_x\rangle$ defines electric quadrupole moment of $B_{2g}$ symmetry localized on four oxygen sites:
\begin{equation}
Q_{xy}=\sum_{i}{\hat Q}_{xy}(i)=Q_{B_{2g}}\,\langle {\hat\sigma}_x\rangle \,.
\end{equation}
It should be emphasized that the quadrupole moment has an electronic orbital origin (see Fig.\,\ref{fig2}) and has nothing to do with any CuO$_4$ plaquette's distortions or charge imbalance between the density of holes at the oxygen sites. It is worth noting that usually a spontaneous imbalance between
the density of holes at the oxygen sites in the unit cell is related to a so-called $nematic$ order.
Order parameter $\langle {\hat\sigma}_y\rangle$ defines  an antiferromagnetic (staggered) ordering of oxygen orbital  moments localized on four oxygen sites:
\begin{equation}
\langle {\hat G}_z\rangle =\langle {\hat l}_{1z}-{\hat l}_{2z}+{\hat l}_{3z}-{\hat l}_{4z}\rangle =g_L\,\langle {\hat\sigma}_y\rangle \, ,
\end{equation}
where $g_L\approx$\,-1.0, if to make use of estimates of the cluster model\,\cite{Moskvin-FNT-11}.  In other words, maximal value of antiferromagnetic order parameter $G_z$ corresponds to a staggered order of unexpectedly large oxygen orbital magnetic moments $m_z\approx 0.25\,\beta_e$. In contrast with the net orbital moment $M_z$ the $G_z$ order cannot be easily quenched by low-symmetry crystal fields.   
Fig.\,\ref{fig2} shows an illustration of the $G_z$ order in the CuO$_4$ plaquette. 
In fact, both quadrupole moment $Q_{B_{2g}}$ and local antiferromagnetic ordering of oxygen orbital moments $G_z$ do result from the ${}^1A_{1g}$-${}^{1}B_{2g}$ mixing effect, in other words, these are a result of the symmetry breaking. It should be emphasized that the 
$G_z$ order resembles the hotly discussed order of circulating currents,  proposed by Varma\,\cite{Varma}. 

Two unconventional vectorial order parameters are associated with the ${}^1A_{1g}$-${}^{3}B_{2g}$ singlet-triplet mixing effect:  $\langle {\hat{\bf V}} {\hat Q}_{xy}\rangle$ and $\langle {\hat{\bf V}} {\hat G}_{z}\rangle$. It should be noted that corresponding orderings do not imply independent $\langle {\hat{\bf V}} \rangle$,  $\langle {\hat Q}_{xy}\rangle$ or $\langle {\hat G}_{z}\rangle$ orders. Moreover, the $\langle {\hat{\bf V}} {\hat Q}_{xy}\rangle$ and $\langle {\hat{\bf V}} {\hat G}_{z}\rangle$ orders imply all the mean values $\langle {\hat{\bf S}} \rangle$, $\langle {\hat{\bf V}} \rangle$, $\langle {\hat Q}_{xy}\rangle$, $\langle {\hat G}_{z}\rangle$ for CuO$_{4}^{5-}$ center together with their
on-site counterparts such as $\langle {\hat{\bf S}}_i \rangle$, $\langle {\hat Q}_{xy}(i)\rangle$, $\langle {\hat l}_{iz} \rangle$ ($i=Cu,\,O_{1,2,3,4}$) turn into zero, at least in first order on the ${}^1A_{1g}$-${}^{3}B_{2g}$ mixing parameters. 
All novel orbital and spin-orbital order parameters appear to be  strongly  hidden, or the hard-to-detect ones and  can be revealed only by specific experimental technique, for instance, by magnetic polarized neutron diffraction. %which is able to discriminate between ${\bf S}$- and ${\bf V}$-type orders and provide an unique opportunity to inspect unconventional spin-quadrupole $\langle {\hat {\bf V}} {\hat Q}_{xy}\rangle$, spin-dipole $\langle {\hat {\bf V}} {\hat d}_{x,y}\rangle$, locally staggered oxygen orbital $\langle {\hat G_z} \rangle$, and oxygen toroidal orbital  $\langle {\hat {\bf T}}\rangle$ orderings, which are determined by the mixing of the ZR singlet $^{1}A_{1g}$ with non-ZR terms $^{3}B_{2g}$, ${}^{3}E_{u}$, $^{1}B_{2g}$, ${}^{1}E_{u}$, respectively. All these  novel spin-orbital and orbital orders break both the time reversal and tetragonal symmetry, however, these can  preserve the translation symmetry of the lattice. 

{\bf Novel orbital modes as seen by neutron diffraction}.
Matrix element of the spin interaction of a neutron with a CuO$_4^{5-}$ center can be written as follows\,\cite{Trammell}:
$$
	\langle SM_S\Gamma\mu|{\hat V}_{{\bf p}{\bf p^{\prime}}}|S^{\prime}M_{S^{\prime}}\Gamma^{\prime}\mu^{\prime}\rangle =-\frac{4\pi \hbar^2}{m}r_0\gamma 
$$
\begin{equation}	
	\langle SM_S\Gamma\mu|\sum_{\nu} {\hat {\bf s}}_{\nu}e^{i{\bf q}{\bf r}_{{\nu}}}|S^{\prime}M_{S^{\prime}}\Gamma^{\prime}\mu^{\prime}\rangle \cdot \left({\bf S}_n-({\bf e}\cdot {\bf S}_n)\cdot {\bf e}\right)\, ,
	\label{ME}
\end{equation}
where sum runs on the two holes ($\nu =1,2$); ${\bf s}_{l\nu}$ is the hole spin;
${\bf e} = \frac{\bf q}{q}$ a unit scattering vector; $r_{0}$  electromagnetic electron radius;  $\gamma = -1.913$  neutron magnetic moment in nuclear Bohr magnetons. Introducing $S$\,-\,$V$ representation we make a replacement for the spin magnetic amplitude:
$$
\sum_{\nu} {\hat {\bf s}}_{\nu}e^{i{\bf q}{\bf r}_{{\nu}}}=\frac{1}{2}{\hat {\bf S}}(e^{i{\bf q}{\bf r}_{1}}+e^{i{\bf q}{\bf r}_{2}})+\frac{1}{2}{\hat {\bf V}}(e^{i{\bf q}{\bf r}_{1}}-e^{i{\bf q}{\bf r}_{2}})
$$
\begin{equation}	
={\hat {\bf S}}{\hat f}_S({\bf q})+{\hat {\bf V}}{\hat f}_V({\bf q})\,.
\label{spin}
\end{equation}
 In other words, we introduce the $S$- and $V$-type spin-orbital operators for CuO$_4^{5-}$ centers with corresponding order parameters which can be detected by polarized neutrons. %Orbital operators ${\hat f}_S$ and ${\hat f}_V$ must transform due to irreducible representation $\gamma$ of the $D_{4h}$ point group:$\gamma \in \Gamma \times \Gamma^{\prime}$.
  As well as spin operator ${\hat {\bf V}}$ changes spin multiplicity, the ${\hat f}_V({\bf q})$ operator changes orbital state.
  For nonzero orbital matrix elements neglecting two-site integrals we obtain
\begin{equation}	
	\langle {}^{1}A_{1g}|{\hat f}_V({\bf q})|{}^{3}B_{2g}\rangle =-\frac{g_L}{4}(\cos q_xl+\cos q_yl)\langle p_x|e^{i{\bf q}{\bf r}}|p_y\rangle \,;
\end{equation} 
$$	
	\langle {}^{1}A_{1g}|{\hat f}_V({\bf q}|{}^{3}E_{ux,y}\rangle =	\mp\frac{ig_L}{2\sqrt{2}}(c_{\pi}\sin q_{y,x}l\langle p_{y,x}|e^{i{\bf q}{\bf r}}|p_{x,y}\rangle 
$$	
\begin{eqnarray}	
-c_{\sigma} \sin q_{x,y}l\langle p_{x,y}|e^{i{\bf q}{\bf r}}|p_{x,y}\rangle )\, ;
\end{eqnarray} 
\begin{equation}	
	\langle p_{x,y}|e^{i{\bf q}{\bf r}}|p_{y,x}\rangle =-3\langle j_2(qr)\rangle_{2p}e_xe_y\,,
\end{equation}
\begin{equation}	
	\langle p_{x,y}|e^{i{\bf q}{\bf r}}|p_{x,y}\rangle =\langle j_0(qr)\rangle_{2p}-\frac{3}{2}\langle j_2(qr)\rangle_{2p}(e_{x,y}^2-e_{y,x}^2)\,,
\end{equation}
where $\langle j_l(qr)\rangle_{2p}$ is a radial average of spherical Bessel function, $l=R_{CuO}=a/2$, the "-" and "+" signs are assigned to matrix elements with $E_{ux}$ and $E_{uy}$, respectively.  It should be noted that at ${\bf q}=0$: ${\hat f}_S(0)=1$, ${\hat f}_V(0)=0$.

Matrix element for the coupling of the neutron spin with the electron orbital moment can be written as (\ref{ME}), if the spin operator (\ref{spin}) to replace by an effective ${\bf q}$-dependent orbital operator as follows\,\cite{Trammell}  
\begin{equation}
{\hat{\bf \Lambda}}({\bf q})=\frac{1}{4}\sum_{n=1}^4\sum_{\nu} \left\{{\hat {\bf l}}_{n\nu},f({\bf q}\cdot{\bf r}_{n\nu})\right\}e^{i{\bf q}{\bf R}_{{n}}}\, ,	
\end{equation}
where ${\hat {\bf l}}_{n\nu}$ is the orbital moment operator for $\nu$-hole on the $n$-th oxygen site, $n$ runs over all four oxygen sites in the CuO$_4^{5-}$ center, $\left\{\,,\,\right\}$ is the anticommutator, 
$$
f({\bf q}\cdot{\bf r})=2\sum_{n=0}^{\infty}\frac{(i{\bf q}\cdot{\bf r})^n}{n!(n+2)}=\sum_{l=0}^{\infty}i^l(2l+1)g_l(qr)P_l(\cos\theta)\,,
$$
where $P_l(\cos\theta)$ is the Legendre polynomial, $\theta$ being the angle between ${\bf q}$ and ${\bf r}$. The functions $g_l(qr)$ are similar to the spherical Bessel functions $j_l(qr)$ which appear in the expansion of $e^{i{\bf q}\cdot{\bf r}}$\,(see Ref.\onlinecite{Trammell} for detail).

\begin{figure}[h]
\includegraphics[width=6.5cm,angle=0]{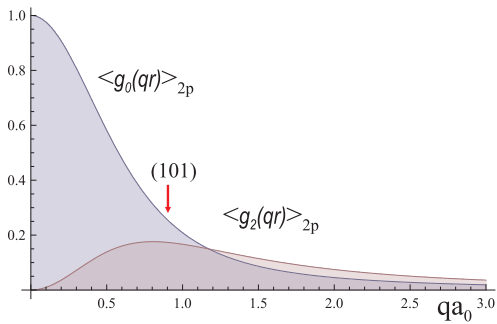}
\caption{q-dependence of $\langle g_{0,2}(qr)\rangle_{2p}$ given hydrogenic 2p functions (a$_0$ is Bohr radius). Arrow point to q-value for Bragg vector (101) in YBa$_2$Cu$_3$O$_{6+x}$.} 
\label{fig3}
\end{figure}

For nonzero orbital matrix elements we obtain after some routine algebra 
$$	
	\langle {}^{1}A_{1g}|{\hat{\bf \Lambda}}({\bf q})|{}^{1}B_{2g}\rangle =
$$	
\begin{equation}	
	\frac{g_L}{8}(\cos q_xl-\cos q_yl)\langle p_x|\left\{{\hat {\bf l}},f({\bf q}\cdot{\bf r})\right\}|p_y\rangle \, ;
\end{equation}
$$	
	\langle {}^{1}A_{1g}|{\hat{\bf \Lambda}}({\bf q})|{}^{1}E_{ux,y}\rangle =
$$
\begin{equation}
\pm\frac{ig_L}{4\sqrt{2}}c_{\pi}\sin q_{y,x}l\langle p_{y,x}|\left\{{\hat {\bf l}},f({\bf q}\cdot{\bf r})\right\}|p_{x,y}\rangle \,,
\label{toroid}
\end{equation}
where $\langle p_x|\left\{{\hat {\bf l}},f({\bf q}\cdot{\bf r})\right\}|p_y\rangle =-\langle p_y|\left\{{\hat {\bf l}},f({\bf q}\cdot{\bf r})\right\}|p_x\rangle $ and 
$$
\langle p_x|\left\{{\hat  l}_z,f({\bf q}\cdot{\bf r})\right\}|p_y\rangle	=
$$
\begin{equation}
-i\left(2\langle g_0(qr)\rangle_{2p} +\langle g_2(qr)\rangle_{2p} (1-3e_z^2)\right)\, ;
\end{equation}
\begin{equation}
\langle p_x|\left\{{\hat  l}_x,f({\bf q}\cdot{\bf r})\right\}|p_y\rangle	=-3i\langle g_2(qr)\rangle_{2p} e_xe_z\, ;
\end{equation}
\begin{equation}
\langle p_x|\left\{{\hat  l}_y,f({\bf q}\cdot{\bf r})\right\}|p_y\rangle	=-3i\langle g_2(qr)\rangle_{2p} e_ye_z \, .
\end{equation}
The "+" and "-" signs in (\ref{toroid}) are assigned to matrix elements with $E_{ux}$ and $E_{uy}$, respectively.
Thus, the orbital vectorial operator ${\hat{\bf \Lambda}}({\bf q})$ on the basis of the non-ZR multiplet can be replaced by an effective operator as follows:
\begin{equation}
{\hat{\bf \Lambda}}({\bf q})={\bf L}_G({\bf q}){\hat G}_z+	{\bf \stackrel{\leftrightarrow}{L}}_T({\bf q}){\hat{\bf T}}\,.
\end{equation}
In contrast with the spin moment the oxygen orbital moment directed perpendicular to the  CuO$_4$ plaquette in the $G_z$ or $T_{x,y}$ modes   induces an effective magnetic coupling both with $z$- and $x$-, and/or $y$-components of the neutron spin. In other words, neutrons see the effective orbital magnetic moments to be tilted in ${\bf c}^*$-${\bf q}$ plane.  Only for ${\bf q}=0$ $\langle g_0(qr)\rangle_{2p}=1$, $\langle g_2(qr)\rangle_{2p}=0$ that is ${\bf L}_G\parallel {\bf z}$, while for nonzero Bragg vectors such as ${\bf q}=(01L)$ or $(10L)$ $\langle g_0(qr)\rangle_{2p}$ and $\langle g_2(qr)\rangle_{2p}$ are of a comparable magnitude\,\cite{Trammell}, that is ${\bf L}_G$ can have sizeable $xy$-plane components. For illustration, in Fig\,\ref{fig3}  we present the 
q-dependence of $\langle g_{0,2}(qr)\rangle_{2p}$ given hydrogenic 2p functions that clearly supports our message. 
It is worth noting one more that this seeming tilting depends both on magnitude and direction of neutron scattering vector ${\bf q}$.

We see that the the magnetic polarized neutron diffraction measurements provide an unique opportunity to inspect unconventional spin-quadrupole $\langle {\hat {\bf V}} {\hat Q}_{xy}\rangle$, spin-dipole $\langle {\hat {\bf V}} {\hat d}_{x,y}\rangle$, locally staggered oxygen orbital $\langle {\hat G_z} \rangle$, and oxygen toroidal orbital  $\langle {\hat {\bf T}}\rangle$ orderings, which are determined by the mixing of the ZR singlet $^{1}A_{1g}$ with non-ZR terms $^{3}B_{2g}$, ${}^{3}E_{u}$, $^{1}B_{2g}$, ${}^{1}E_{u}$, respectively. Exps.(5)-(8) and (10)-(14) provide ${\bf q}$-dependence of the formfactors for respective neutron scattering.

All these  novel orbital and  spin-orbital orders break both the time reversal and tetragonal (Z$_4$) symmetry, however, 
our analysis does show that polarized elastic neutron scattering measurements performed  at the Bragg scattering wave vectors can detect two hidden modes which preserve translational symmetry of the lattice, these are the spin-quadrupole $\langle {\hat {\bf V}} {\hat Q}_{xy}\rangle$-mode at ${\bf q}=(11L)$ (see Exps. (5) and (7)) or locally staggered orbital $G_z$-mode at the Bragg vectors such as ${\bf q}=(01L)$ or $(10L)$ (see Exps. (10) and (12)-(14)).     
Namely the latter type of the long range magnetic order has been experimentally observed  in the pseudogap phase for three different cuprate families, YBCO,  Hg1201\,\cite{Bourges}, and Bi2212\,\cite{Bi2212}. Similar short range bidimensional order while occuring at a temperature far below T$^*$ has been  observed in LSCO system\,\cite{Bourges}.

Furthermore, to explain the experimental data\,\cite{Bourges,Bi2212} we do not need
to engage the spin-orbital coupling,  quantum corrections\,\cite{Varma} or
orbital currents involving the apical oxygens\,\cite{Weber} as the measured polarization effects can be explained with the  locally staggered oxygen orbital moments orthogonal to the CuO$_2$ planes. It is worth noting that the $G_z$-type ordering preserving the translational symmetry cannot be detected in the polarized elastic neutron scattering measurements performed  at the Bragg scattering wave vector such as ${\bf q}=(11L)$ that does explain earlier unsuccessful polarized neutron reports\,\cite{Lee}.

Oxygen orbital moments must inevitably generate local magnetic fields, first of all it concerns a giant $\sim$ 1\,T field (given oxygen magnetic moment of $\sim$\,0.1\,$\mu_B$) at the oxygen nuclei directed perpendicular to the CuO$_2$ plane. 
However, the $^{17}$O NMR data on very different cuprates \,\cite{17O-NMR} do not reveal signatures of static G$_z$ type
mode. At present, there are no published $^{63,65}$Cu or $^{17}$O NMR studies which give clear results concerning the existence or absence of fields of the predicted magnitude in YBCO, La-214, Hg1201 or Bi2212. 
The $G_z$-type orbital magnetic order, as any other moment patterns which have reflection symmetry across the Cu-O-Cu bonds would generate a zero magnetic field on yttrium and barium sites in YBa$_2$Cu$_3$O$_{6+\delta}$, YBa$_2$Cu$_4$O$_{8}$, Y$_2$Ba$_4$Cu$_7$O$_{15-\delta}$,  thus making direct $^{89}$Y and $^{135,137}$Ba NQR/NMR  methods as "silent local probes"\, despite their pronounced sensitivity for weak local magnetic fields. This reconciles  the "non-observance"\, results obtained by ${}^{89}$Y NMR in superconducting Y$_2$Ba$_4$Cu$_7$O$_{15-\delta}$ and   
  ${}^{135,137}$Ba NQR in superconducting YBa$_2$Cu$_4$O$_{8}$\,\cite{Strassle} with neutron scattering results\,\cite{Bourges}.
The ZF-$\mu$SR measurements in  YBa$_2$Cu$_3$O$_{6+\delta}$ and La$_{2-x}$Sr$_x$CuO$_4$\,\cite{Sonier} have also found no evidence for the onset of magnetic order at
the pseudogap temperature T$^*$. The NMR and $\mu$SR experiments clearly rule  static G$_z$ type order out. 
The failure to detect orbital-like magnetic order of the kind observed by spin-polarized neutron diffraction surely indicates that the local fields are rapidly fluctuating outside the $\mu$SR or NMR time window or the order is associated with a small minority phase that evolves with hole doping\,\cite{Sonier}.

%It is worth noting that our model conclusions have much in common with predictions of the circulating current theory of the pseudogap state\,\cite{Varma}, however, ...

%Our analysis\,\cite{Moskvin-order parameters} does show that polarized elastic neutron scattering measurements performed  at the Bragg scattering wave vectors can detect only two hidden modes which preserve translational symmetry of the lattice, the spin-quadrupole $\langle {\hat {\bf V}} {\hat Q}_{xy}\rangle$-mode at ${\bf q}=(11L)$ or locally staggered orbital $G_z$-mode at the Bragg vectors such as ${\bf q}=(01L)$ or $(10L)$. Namely the latter type of the long range magnetic order has been experimentally observed  in a pseudogap phase for two different cuprate families, YBCO  and Hg1201\,\cite{Bourges}. Similar short range bidimensional order has been  observed in LSCO system\,\cite{Bourges}. Interestingly that in contrast with the spin moment the oxygen orbital moment directed perpendicular to the  CuO$_4$ plaquette in the $G_z$ or $T_{x,y}$ phases   induces an effective magnetic coupling both with $z$- and $x$-, and/or $y$-   components of the neutron spin. In other words, the neutron sees the effective orbital magnetic moments to be tilted in ${\bf c}^*$-${\bf q}$ plane. The seeming tilting depends both on direction and magnitude of neutron scattering vector ${\bf q}$.

{\bf Summary}. Instead of inactive ZR singlet the ground state of the hole CuO$_4^{5-}$ center  should be described by a complex $^1A_{1g}$-$^{1,3}B_{2g}$-$^{1,3}E_u$ valent multiplet with several unconventional hidden order parameters such as  intra-plaquette's staggered order of Ising-like oxygen orbital magnetic moments or a complex spin-quadrupole ordering.
Non-ZR hole centers are believed to be an essential ingredient of both the hole and electron doped cuprates\,\cite{Moskvin-FNT-07}. In particular, these specify many features of the pseudogap regime (see, e.g., Refs.\,\onlinecite{Panagopoulos,Kerr}) which indeed manifests clearly seen competing orders.  
The most convincing evidence of the non-ZR hole centers in cuprates was obtained by experimental findings\,\cite{Bourges} which provide the first direct evidence of a hidden order incompatible with simple
ZR singlet. As we argue, firstly one needs to reconsider the role of oxygen orbitals, especially the in-plane
p-orbitals. Indeed, only the competition of the $\sigma$ and $\pi$ oxygen holes can explain the emergence of oxygen orbital moments. 
It is worth noting that the multiplet structure of the ground state makes the hole [CuO$_4$]$^{5-}$ centers in cuprates to be the pseudo-Jahn-Teller centers\,\cite{PJT}. Thus, the non-ZR hole centers can be responsible   for different magnetic and lattice effects which are often addressed to be signatures of the magnetic (spin-fluctuation) or electron-phonon mechanism of the high-temperature superconductivity.

We believe that a large body of puzzling effects governed by the non-ZR structure of the hole centers are secondary ones and are not of primary importance for the high-T$_c$ superconductivity. However, we need to understand these phenomena to understand high-T$_c$ puzzle itself.
A full exhaustive explanation of all the cuprate physics governed by the non-ZR hole centers cannot be given at this point, but  we propose a consistent picture that can  be successfully used for the distinctive and complex description of today and future experimental findings.

The work was partially supported by RFBR grants Nos. 10-02-96032 and 12-02-01039.

\end{document}